\documentclass[aps,prd,twocolumn,superscriptaddress]{revtex4-1}
\usepackage{color}
\usepackage[pdftex]{graphicx}
 \usepackage{hyperref}

\def\beq{\begin{equation}}
\def\eeq{\end{equation}}
\def\bea{\begin{eqnarray}}
\def\eea{\end{eqnarray}}

\def\bwt{\begin{widetext}}
\def\ewt{\end{widetext}}
\usepackage{amssym,aas_macros}

\begin{document}

\title{Neutron Stars and the Cosmological Constant Problem}

\author{Farbod Kamiab}\email{fkamiab@perimeterinstitute.ca}
\affiliation{Department of Physics and Astronomy, University of Waterloo, Waterloo, ON, N2L 3G1, Canada}
\affiliation{Perimeter Institute
for Theoretical Physics, 31 Caroline St. N., Waterloo, ON, N2L 2Y5, Canada}

\author{Niayesh Afshordi}\email{nafshordi@perimeterinstitute.ca}
\affiliation{Perimeter Institute
for Theoretical Physics, 31 Caroline St. N., Waterloo, ON, N2L 2Y5, Canada}
\affiliation{Department of Physics and Astronomy, University of Waterloo, Waterloo, ON, N2L 3G1, Canada}

\date{\today}
\preprint{astro-ph/yymmnnn}

\begin{abstract}
The gravitational aether theory is a modification of general relativity that decouples vacuum energy from gravity, and thus can potentially address the cosmological constant problem. The classical theory is distinguishable from general relativity only in the presence of relativistic pressure (or vorticity). Since the interior of neutron stars has high pressure and as their mass and radius can be measured observationally, they are the perfect laboratory for testing the validity of the aether theory. In this paper, we solve the equations of stellar structure for the gravitational aether theory and find the predicted mass-radius relation of non-rotating neutron stars using two different realistic proposals for the equation of state of nuclear matter. We find that the maximum neutron star mass predicted by the aether theory is 12\% - 16\% less than the maximum mass predicted by general relativity assuming these two equations of state. We also show that the effect of aether is similar to modifying the equation of state in general relativity. The effective pressure of the neutron star given by the aether theory at a fiducial density differs from the values given by the two nuclear equations of state to an extent that can be constrained using future gravitational wave observations of neutron stars in compact systems. This is a promising way to test the aether theory if further progress is made in constraining the equation of state of nuclear matter in densities above the nuclear saturation density. 

\end{abstract}

\maketitle
\section{Introduction}

\label{sec:intro}

One of the fundamental challenges of modern physics is to solve the cosmological constant problem. This problem, in its various forms, has been with us since the beginning of the 20th century. At the end of the century, data from Type Ia supernovae pointed to the fact that the expansion of the universe is accelerating \cite{riess, garnavich, perlmutter}. A constant is needed in the general relativistic description of gravity in order to explain the acceleration of the cosmic expansion. One can interpret this ``cosmological constant" as the energy of the vacuum (popularized as dark energy). Quantum field theory predicts a value for this energy. However the predicted value is much larger than the value found from observational data. This is the so-called ``old cosmological constant problem" \cite{weinberg}.  The new problem is why this constant is very small but not zero. Another problem is that the values of dark energy density and matter density are found to be of the same order at the present epoch oddly suggesting that we are living in a special time in the history of the universe. This is the so-called ``coincidence problem". 

One possible solution to the cosmological constant problems was suggested by one of us in \cite{afshordi1}, which modifies the Einstein equation in the following way so that the vacuum does not gravitate:
\beq
(8\pi \mathcal{G})^{-1} G_{\mu\nu}[g_{\mu\nu}] = T_{\mu\nu} - \frac{1}{4}Tg_{\mu\nu} + ...~ ,
\eeq \label{aether1}

\noindent where $\mathcal{G}=4G_N/3$ and $G_N$ is the usual Newton's gravitational constant. By subtracting the trace of the energy momentum tensor on the right-hand side, the Einstein equation becomes insensitive to the vacuum energy density, $\rho_{\rm vac}$, where $T_{\mu\nu} = \rho_{\rm vac}g_{\mu\nu} + {\rm excitations}$. As energy and momentum are conserved, the divergence of $T_{\mu\nu}$ vanishes. By definition the divergence of $G_{\mu\nu}$ also vanishes through the Bianchi identities. Therefore, if we want to subtract the trace of the energy momentum tensor, we need to add a suitable term to it so that the divergence of the right-hand side of Eq.\  \ref{aether1} vanishes. It was suggested in \cite{afshordi1} that this term can be the energy momentum tensor of a perfect fluid $\mathcal{T}_{\mu\nu}$, which is dubbed ``gravitational aether". With this term, Eq.\  \ref{aether1} takes the form:
\beq
(8\pi \mathcal{G})^{-1} G_{\mu\nu}[g_{\mu\nu}] = T_{\mu\nu} - \frac{1}{4}Tg_{\mu\nu} + \mathcal{T}_{\mu\nu},\label{aether}
\eeq

\beq
\mathcal{T}_{\mu\nu} = \mathcal{P}(\mathcal{U}_\mu \mathcal{U}_\nu +g_{\mu\nu}),
\eeq

\noindent where $\mathcal{P}$ and $\mathcal{U}_\mu$ are the pressure and four-velocity of the gravitational aether. For the right hand side of Eq.\  \ref{aether} to be divergenceless, we require:
\beq
{{{\mathcal{T}}_{\mu}}^{\nu}}_{;\nu} =\frac{1}{4} T_{,\mu} .\label{aether_cont} \label{1}
\eeq

It is argued in \cite{afshordi1} that the pressure and four-velocity of the gravitational aether are dynamically fixed in terms of $T_{\mu\nu}$ via Eq.\  \ref{1}. Furthermore, it is consistent with all the current precision tests of gravity and cosmological observations \cite{siavash}. 

With the metric being blind to vacuum energy, the gravitational aether theory solves the old cosmological constant problem.  But how does it explain the coincidence problem? In \cite{chanda} the authors study static black hole solutions in the gravitational aether and argue that aether couples the spacetime metric close to the black hole horizon, to the metric at infinity. They then show that this can lead to an accelerating cosmological solution, far from the horizon. This connection between the formation of stellar black holes and the acceleration of the expansion of the universe can potentially solve the coincidence problem, through Planck-suppressed corrections in black hole physics. 

All this means that the gravitational aether theory is an attractive alternative to general relativity. However, the theory is good only if it can make definite predictions that are consistent with current and future observational results. In astrophysical situations where gravity due to vorticity or  pressure is negligible, the effects of the aether theory are indistinguishable from general relativity \cite{afshordi1, siavash}. The aether theory can be tested only in high pressures and strong gravitational forces. The interior of neutron stars satisfies these conditions. In addition pulsars can be observed and studied empirically,  enabling observational tests of theoretical models. 

Two sets of models define the structure of neutron stars. First models describing gravity that is the binding force of the star, and second models describing the elementary constituents at the core of neutron stars and their repulsive forces that work against gravity and prevent the neutron star from collapsing and forming a black hole.  Apart from quantum degeneracy pressure, strong nuclear interactions are the main sources of pressure inside neutron stars. Various nuclear models give different pressure-density relations (equations of state) for the interior of neutron stars. Much of the uncertainty in the study of neutron stars is due to the lack of knowledge of the correct equation of state (EOS). Having this equation and using a description of gravity we can find the mass-radius relation of neutron stars in static equilibrium. In other words, for each neutron star of a given radius, we can find the mass for which the repulsive and attractive forces cancel and ensure the hydrostatic equilibrium of the star. In the context of general relativity (or the gravitational aether theory), in contrast with Newtonian physics, this mass-radius relation has a maximum mass ${\rm M}_{\rm max}$ beyond which no neutron star would exist and only black holes could have higher masses \cite{lattimer2007}. Therefore observations of high-mass neutron stars have the potential to constrain some equations of state and rule out others.  Such an observation was made recently \cite{demorest}. A millisecond pulsar was observed and its mass (1.97 $\pm$ 0.04 M$_\odot$) was calculated using the Shapiro delay of the pulsar light due to its companion, a half solar-mass white dwarf.  The high mass of this pulsar provides a lower limit on the maximum mass of neutron stars and rules out a number of proposed equations of state \cite{ozel&demorest}. Another observation was reported in \cite{blackwidow}. The authors presented evidence that the black widow pulsar, PSR B1957+20, has a high mass. Their best fit pulsar mass was $\sim$ 2.40 $\pm$ 0.12 M$_\odot$. A number of assumptions in the theoretical modelling of the pulsar contributed to the uncertainty in this number. Considering different constraints, the authors inferred a lower limit to the pulsar mass of M $>$ 1.66 M$_\odot$. Future observations of neutron stars will put additional constraints on the EOS.

In this paper, we calculate the mass-radius relation predicted by the gravitational aether theory for two well-known equations of state. The first EOS (hereafter denoted FPS) was calculated by Friedman and Pandharipande and improved by Lorenz, Ravenhall and Pethick \cite{fps, lorenz}. This equation of state is based on variational calculations over a wide density range using a realistic nuclear hamiltonian that contains two- and three-nucleon interactions, and fits the nucleon-nucleon scattering, as well as nuclear matter data. The Skyrme model is used in the FPS equation of state. In this model, the effective interaction has the spatial character of a two-body delta function plus derivatives. The second EOS was calculated by Akmal and Pandharipande (hereafter denoted AP3)\cite{ap}. Some improvements of this calculation compared to FPS are the use of Green�s function Monte Carlo (GFMC) methods in the variational theory and including two-pion exchange three-nucleon interaction and isospin symmetry breaking terms in the hamiltonian. \"Ozel and Psaltis have shown in \cite{ozel} that the complete mass-radius relation of neutron stars can be reproduced to high accuracy for all proposed equations of state, when the pressure of the neutron star is specified at three fiducial densities beyond the nuclear saturation density of $\rho_{\rm ns}\sim 2.7\times 10^{14}$~g~cm~$^{-3}$. As they have calculated these values of pressure for the FPS and AP3 equations of state, we will use their method to reproduce these two equations of state for densities higher than $\rho_{0}$ which is a parameter to be adjusted for each EOS. For densities below $\rho_{0}$ (the outer layers of the neutron star) we will use the SLy (Skyrme
Lyon) equation of state calculated by Douchin and Haensel in \cite{douchin}. This equation is based on the effective nuclear interaction SLy of the Skyrme type, which is useful in describing the properties of very neutron rich matter.

The structure of this paper is as follows: In section \ref{cosmo}, we summarize the phenomenological implications of the gravitational aether theory studied in \cite{siavash}. In section \ref{sec:aether}, we derive the equations of stellar structure for the gravitational aether theory and relate the mass predicted by the theory to the observable mass of neutron stars. In section \ref{sec:polytropic}, we solve the equations of stellar structure for a simplistic polytropic equation of state and explain the numerical method used. In section \ref{sec:realistic}, we solve the equations for the realistic FPS and AP3 equations of state, find the mass-radius relation of neutron stars predicted by the gravitational aether theory and compare it to the prediction of general relativity. An equivalent description of the problem in terms of a modified EOS will be described in section \ref{sec:aetherEOS}. Section \ref{sec:conclusion} will include a discussion and the summary of our results.

\section{Phenomenological Implications of the Aether Theory}
\label{cosmo}
In \cite{siavash}, it was shown that the deviations of the aether theory from general relativity can only be significant in situations with relativistic pressure, or (potentially) relativistic vorticity. Furthermore, the authors demonstrate that the theory is consistent with all the current precision tests of gravity and cosmological observations. Here, we summarize their results:

They show that for a perfect fluid with linear equation of state ($p \propto \rho$), the solutions to the gravitational aether theory are identical to those of general relativity only with a renormalized gravitational constant. As the gravitational coupling is not a constant in the aether, the authors find that in the case of homogeneous FLRW cosmology, radiation energy gravitates more strongly than non-relativistic matter. The aether theory implies that gravity should be 33\% stronger in the cosmological radiation era than the predictions of general relativity. 

As the increase of the gravitational constant at around the $T= {\cal O}(1)$ MeV epoch induces an earlier freezeout of the neutron to proton ratio because of a speed-up effect of the increased cosmic expansion, the abundance of $^{4}$He increases sensitively, and the abundance of deuterium (D) increases mildly while the abundance of $^{7}$Be decreases. Comparing the theoretical prediction with the observational light element abundances, the authors found that every light element abundance agrees with the gravitational aether theory within 2$\sigma$. They found notably that $^7$Li fits the data better in the gravitational aether than in the standard big bang nucleosynthesis (which over-predicts $^7$Li abundance by 4-5$\sigma$ \cite{lithium}). The main discrepancy found was with deuterium abundance observed in quasar absorption lines. 

Interestingly, cosmological observations, such as the CMB (Cosmic Microwave Background) \cite{atakama} and the Ly-$\alpha$ forest \cite{seljak} prefer the aether prediction of stronger gravity in the radiation era, which is often interpreted as a larger effective number of neutrinos. 

The authors in \cite{siavash} also examined the implications for precision tests of gravity using the PPN (parametrized post-Newtonian) formalism \cite{ppn}, and showed that the only PPN parameter that deviates from its general relativistic value is $\zeta_4$, the anomalous coupling to pressure (=1/3 for the aether and 0 for general relativity), that has never been tested experimentally. Finally, they argued that current tests of Earth's gravitomagnetic effect mildly prefer a co-rotation of aether with matter, although they are consistent with an irrotational aether at $2\sigma$ level.

In the current paper, we study the impact of the anomalous coupling to pressure in the aether on the structure of neutron stars and therefore suggest a novel way of putting constraints on the value of $\zeta_4$ using mass measurements of neutron stars. We will return to this point in section \ref{sec:realistic}.

\section{The aether equations of stellar structure}

\label{sec:aether} 

As we assume a spherically symmetric static star, the metric will take the form:

\[ g_{\mu \nu} = \left( \begin{array}{cccc}
-B(r) & 0 & 0 & 0 \\
0 & A(r) & 0 & 0\\
0 & 0 & r^2  & 0\\
0 & 0 & 0 & r^2 \sin^2 \theta
\end{array} \right),\]
\noindent with the line element being $ds^2=g_{\mu \nu}dx^\mu dx^\nu$. With this metric and since the problem is static the energy momentum tensor will be:

\[ T^\mu_\nu = \left( \begin{array}{cccc}
-\epsilon(r) & 0 & 0 & 0 \\
0 & p(r) & 0 & 0\\
0 & 0 & p(r)  & 0\\
0 & 0 & 0 &  p(r)
\end{array} \right),\]
where $\epsilon(r)$ and $p(r)$ are the energy density and pressure at radius  $r$ of the star. The modified Einstein equation has the form:
\beq
(8\pi {G_N})^{-1} G_{\mu\nu}[g_{\mu\nu}] = \tilde{T}_{\mu \nu} \label{modein},
\eeq
\noindent where  $\tilde{T}_{\mu \nu}$ is given by:
\beq
\tilde{T}_{\mu \nu}= (4/3) \big[T_{\mu\nu} - \frac{1}{4}T^\alpha_\alpha g_{\mu\nu} + \mathcal{P}(\mathcal{U}_\mu \mathcal{U}_\nu +g_{\mu\nu})\big],
\eeq

\noindent $T^\alpha_\alpha=3p(r)-\epsilon(r)$ is the trace of the energy momentum tensor and $\mathcal{P}$ and $\mathcal{U}_\mu$ are the pressure and four-velocity of the gravitational aether.  Having $\tilde{T}^{\mu}_{\nu}=g^{\mu\beta}\tilde{T}_{\beta \nu}$, where we have the Einstein summation over index $\beta$, and imposing spherical static conditions on the aether we will get:
\[ \tilde{T}^\mu_\nu = \left( \begin{array}{cccc}
-\tilde{\epsilon}(r) & 0 & 0 & 0 \\
0 & \tilde{p}(r) & 0 & 0\\
0 & 0 & \tilde{p}(r) & 0\\
0 & 0 & 0 &  \tilde{p}(r)
\end{array} \right),\] 
where:
\beq
\begin{array}{rcl} \tilde{\epsilon}(r) & = &     \epsilon(r) + p(r)   ,\\ 
\noalign{\medskip}
\tilde{p}(r) & = & {(1/3)} \ \big[  \ \epsilon(r) + p(r) \ \big]  + (4/3) \ \mathcal{P}(r).  \end{array} \label{densityandpressure}
\eeq

As Eq.\  \ref{modein} is similar to the Einstein equation only with different energy density $\tilde{\epsilon}(r)$ and pressure $\tilde{p}(r)$ given by Eq.\  \ref{densityandpressure}, the equations of stellar structure will be the same as in general relativity only with these updated quantities ($\tilde{\epsilon}$ and $\tilde{p}$). Using the Ricci tensor components and writing the different spherical components of the modified Einstein equation (Eq.\  \ref{modein}) we will have \cite{weinbergbook}:
\beq
R_{rr}= {B'' \over 2B} - {B' \over 4B} \Big({A' \over A}+ {B' \over B}\Big) - {A' \over rA}= -4 \pi{G_N} \big(\tilde{\epsilon}-\tilde{p }\big)A,
\eeq
\beq
R_{\theta \theta}= -1 + {r \over 2A} \Big(-{A' \over A}+ {B' \over B}\Big) + {1 \over A}= -4 \pi{G_N} \big(\tilde{\epsilon}-\tilde{p }\big)r^2,
\eeq
\beq
R_{t t}=  -{B'' \over 2A} + {B' \over 4A} \Big({A' \over A}+ {B' \over B}\Big)- {B' \over rA}=-4 \pi{G_N} \big(\tilde{\epsilon}+3\tilde{p }\big)B,
\eeq
\noindent where the prime superscript denotes the derivative with respect to radius. We have omitted the $R_{\phi \phi}$ equation as it is identical to $R_{\theta \theta}$ because of the spherical symmetry. Rewriting these equations using Eq.\  \ref{densityandpressure} we get:
\beq
{B'' \over 2B} - {B' \over 4B} \Big({A' \over A}+ {B' \over B}\Big) - {A' \over rA}= -{8 \pi \over 3}{G_N} \big({\epsilon+p } - 2\mathcal{P}\big)A \label{eq1},
\eeq
\beq
-1 + {r \over 2A} \Big(-{A' \over A}+ {B' \over B}\Big) + {1 \over A}= -{8 \pi \over 3}{G_N} \big({\epsilon+p } - 2\mathcal{P}\big)r^2 \label{eq2},
\eeq
\beq
-{B'' \over 2A} + {B' \over 4A} \Big({A' \over A}+ {B' \over B}\Big)- {B' \over rA}=-8 \pi{G_N}( \epsilon + p+ 2 \mathcal{P})B.
\eeq

Given our metric, the equation of hydrostatic equilibrium for $p$ and $\epsilon$ is \cite{weinbergbook}: 
\beq
{B' \over B} = - {2 {p}'\over {\epsilon} + {p}} \label{eq3}.
\eeq
In addition to this, the same equation holds for our updated $\tilde{\epsilon}$ and $\tilde{p}$. This equation is not independent from the modified Einstein equations and can be derived from them: 
\beq
{B' \over B} = - {2 \tilde{p}'\over \tilde{\epsilon} + \tilde{p}} = -2 { (\epsilon' +p')/4+\mathcal{P}'\over \epsilon +p + \mathcal{P}} \label{eq4}.
\eeq

Given the suitable boundary conditions, Equations \ref{eq1}, \ref{eq2}, \ref{eq3} and \ref{eq4} along with an equation of state giving $\epsilon(p)$ (the energy density of the star as a function of its pressure) are enough to find our unknowns: $A(r)$, $B(r)$, $p(r)$, $\epsilon(r)$ and $\mathcal{P}(r)$. In practice, this needs to be done numerically.

According to \cite{chanda} the pressure in the vacuum does not vanish and is comparable to the pressure associated with dark energy. This pressure will be negligible for the calculations of neutron star structure. Therefore we can assume that pressure and energy density both vanish outside the neutron star and the metric becomes the familiar Schwarzschild metric for which at $r\geq {\rm R}$:
\beq
B(r)=A^{-1}(r)=1-{2{G_N}{\rm M}({\rm R}) \over r}, \label{schwar}
\eeq

\vspace*{0mm}
\rm 
\noindent where ${\rm R}$ is the radius of the neutron star and ${\rm M}$ is the observed mass of the star given by the aether theory:
\beq
{\rm M}({\rm R}) = \int_0^{\rm R} \tilde{\epsilon}(r)r^2drd\Omega.
\eeq

Using Eq.\  \ref{densityandpressure} this gives:
\beq
{\rm M}({\rm R}) = \int_0^{\rm R}   \big[   \epsilon(r) + p(r)  \big] r^2drd\Omega \label{mobsint}
.
\eeq 
The purpose of this work is to find the ${\rm M}-\rm R$ relation for neutron stars assuming different equations of state.

\section{Numerical solutions for a polytropic equation of state}

\label{sec:polytropic} 

We start by solving the equations of stellar structure for a simple EOS. The polyropic equation of state is a power-law relation between pressure and matter density:
\beq
p=K\rho^\Gamma \label{poly1}.
\eeq

\begin{figure}[t]
\centering
   \includegraphics[scale=0.4]{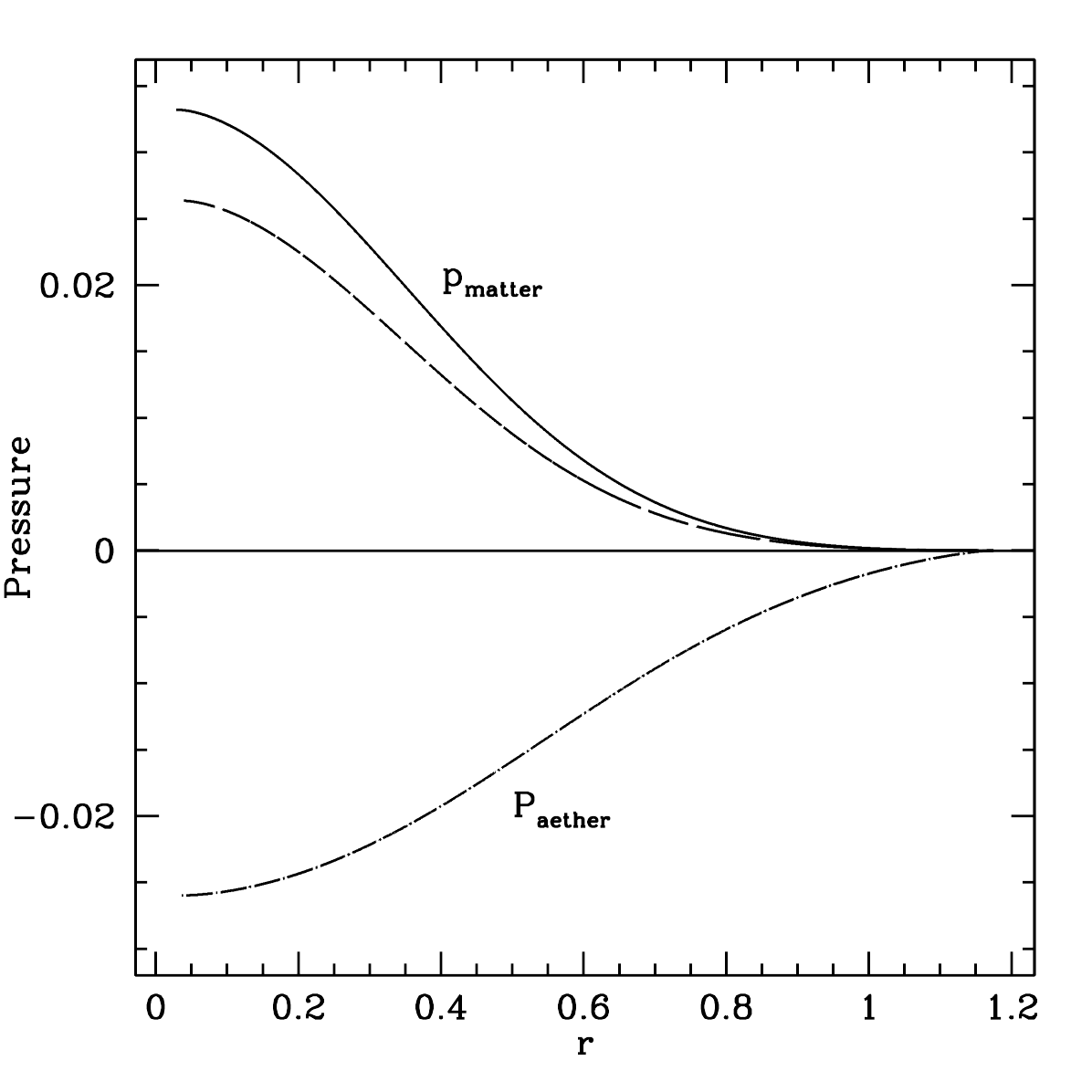}
\caption{The $p(r)-r$ relation of a neutron star of radius $\rm R=1.175$ with a polytropic equation of state $p=K\rho^\Gamma$ where $\Gamma=9/5$ for general relativity (solid) and the aether theory (dashed). The pressure of the aether $\mathcal{P}(r)$ which is negative is shown as well (dashed-dotted). The units have been chosen so that $G_N=1$, $c=1$ and $K=1$.}
\label{fig:p-r}
\end{figure}


The energy density is given as a function of matter density by:
\beq
\epsilon(\rho)=\rho c^2 + \rho \int^\rho_0 {p(\rho')d\rho' \over \rho'^2},
\eeq
where the second term is negligible for non-relativistic matter. For neutron stars this term needs to be taken into account. If we use Eq.\  \ref{poly1} to perform the integration in the second term we will get:
\beq
\epsilon(\rho)=\rho c^2 + {K \rho^\Gamma \over \Gamma-1}.
\eeq

If we choose our units so that $c=1$ and $G_N=1$ (these will fix our time and mass units given a length unit), the energy density as a function of pressure will be:
\beq
\epsilon(p)=({p \over K})^{1 \over \Gamma} + {p \over \Gamma-1} \label{poly2}.
\eeq

The differential equations \ref{eq1}, \ref{eq2}, \ref{eq3} and \ref{eq4} along with Eq.\  \ref{poly2} need to be solved numerically. The boundary conditions are the values of  $A({\rm R})$, $B({\rm R})$, $p({\rm R})$, $\epsilon({\rm R})$ and $\mathcal{P}({\rm R})$ (R is the radius of the neutron star). We set the values of pressure, energy density and aether pressure equal to zero at R:
\beq
\begin{array}{rcl} \epsilon({\rm R}) & = & 0,\\ 
\noalign{\medskip}
 p({\rm R}) & = & 0,\\  
 \noalign{\medskip}
 \mathcal{P}({\rm R}) & = & 0.  \end{array} 
\eeq

The reason why we set the pressure of the aether equal to zero at the boundary of the star is that if the pressure of the aether in the vacuum is very small at infinity it will remain small up to the boundary of the star. This can be understood by writing equation \ref{eq4} in the vacuum:
\beq
{B' \over B} = -2 { \mathcal{P}'\over \mathcal{P}} \label{eq444},
\eeq
which gives:
\beq
\mathcal{P}={\mathcal{P}_{\infty}\over\sqrt{1-2GM/r}}.
\eeq
If $\mathcal{P}_\infty \rightarrow 0$, then $\mathcal{P} \rightarrow 0$ at the boundary of the star. It is straight-forward to generalize this argument to dynamical situations, i.e. aether pressure vanishes in vacuum everywhere if it vanishes at large distances. Therefore, the aether does not affect the binary mass measurements, enabling us to compare our results with the current observations. 

The values of $A({\rm R})$ and $B({\rm R})$ are given by:
\beq
B({\rm R})=A^{-1}({\rm R})=1-{2{\rm M}({\rm R}) \over {\rm R}}.
\eeq
As we do not have the value of ${\rm M}({\rm R})$ and finding it is the purpose of this integration, we will use a shooting method in which for a given radius $\rm R$ we solve the differential equations with different values of ${\rm M}({\rm R})$ starting from $\rm R/2$ to smaller values. For each value of ${\rm M}({\rm R})$ solving the equations gives the energy density and pressure as a function of radius. Using Eq.\  \ref{mobsint} we can find the integrated mass ${\mathcal{ M}}$. The value of ${\rm M}({\rm R})$ for which ${\rm M}({\rm R})={\mathcal{ M}}$ is the correct mass of the neutron star of radius R. For this value the functions $A(r)$, $B(r)$, $p(r)$, $\epsilon(r)$ and $\mathcal{P}(r)$ are well-behaved and correspond to the solutions of equations  \ref{eq1}, \ref{eq2}, \ref{eq3}, \ref{eq4} and \ref{poly2}. For instance, the pressure $p(r)$ given by general relativity and the aether theory is shown in Figure \ref{fig:p-r} for a neutron star of fixed radius. The integration method used is a fourth-order Runge-Kutta. 
\begin{figure}[t]
\centering
   \includegraphics[scale=0.4]{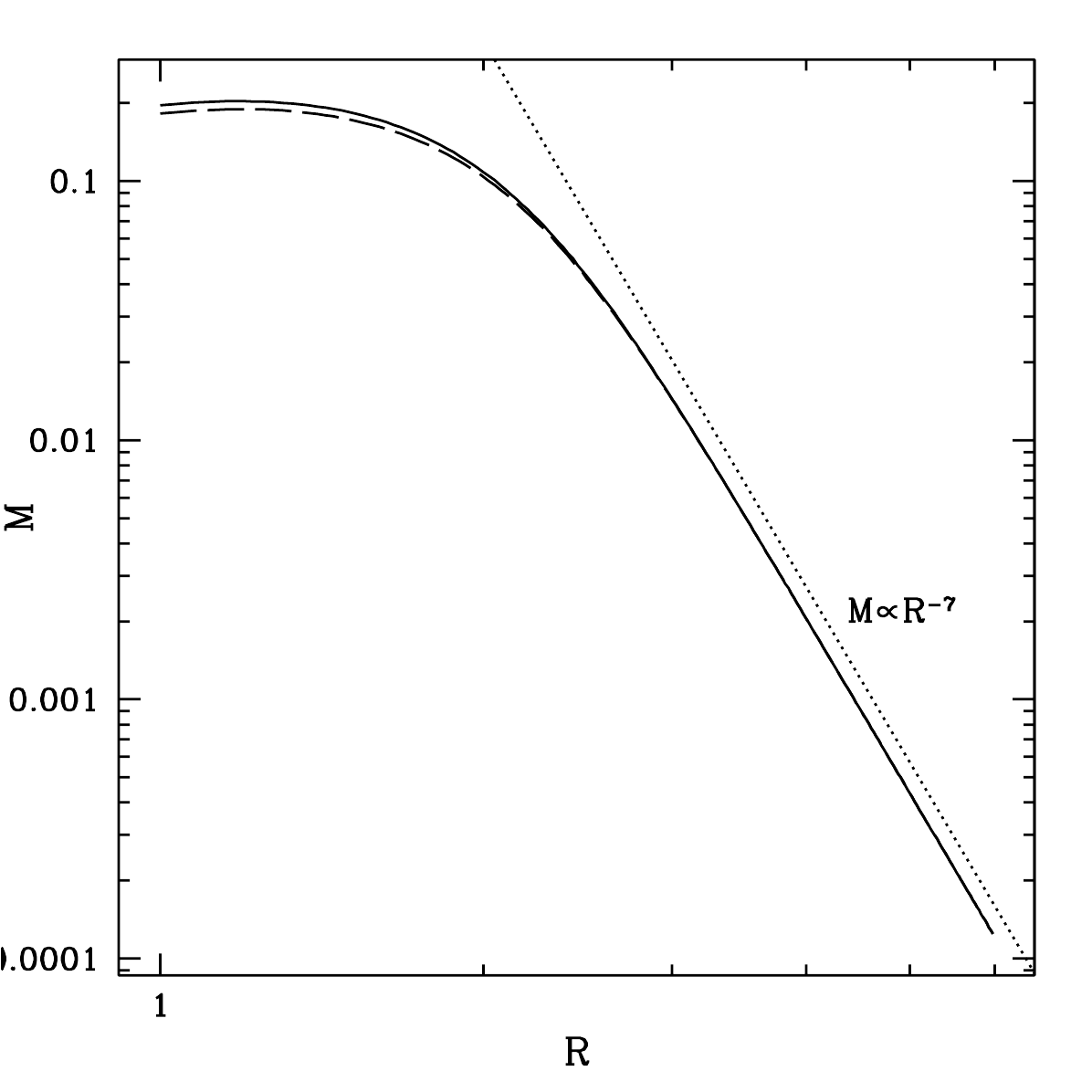}
\caption{The ${\rm M}({\rm R})$ - R relation of a neutron star with a polytropic equation of state $p=K\rho^\Gamma$ where $\Gamma=9/5$ for general relativity (solid) and the aether theory (dashed). The units have been chosen so that $G_N=1$, $c=1$ and $K=1$. We see that the maximum observed mass predicted by the aether theory is less than the maximum mass predicted by general relativity. We also see that in the Newtonian limit (large R) ${\rm M}({\rm R})\propto {\rm R}^{-7}$.}
\label{fig:logMRpoly}
\end{figure}
Figure \ref{fig:logMRpoly} shows the ${\rm M}({\rm R})$ - R relation for neutron stars with a polytropic equation of state $p=K\rho^\Gamma$ with $\Gamma=9/5$ for general relativity and the aether theory in units for which $G_N=1$, $c=1$ and $K=1$. We have chosen $\Gamma=9/5$ as it is consistent with the constraints found by \"Ozel and Psaltis in \cite{ozel} and used in \cite{cooney} to study the structure of neutron stars in $f$(R) gravity theories with perturbative constraints. We see that the aether theory gives a smaller mass for a neutron star of a given radius. This was expected as gravity is stronger in the gravitational aether theory in the relativistic regime, reflected in the $\mathcal{G}=4G_N/3$ relation. Therefore a neutron star of a given radius needs less mass to sustain its hydrostatic equilibrium compared to general relativity.

To understand the behaviour of  ${\rm M}({\rm R})$ in large radii (the Newtonian limit) we can look at the equation of hydrostatic equilibrium in this limit:
\beq
\rho(r) g(r) = {d p(r) \over d r},
\eeq
where $g$ is the gravitational force and $\rho(r) \sim \epsilon(r)$ in the Newtonian limit (as we have set $c=1$). As $p=K\rho^{\Gamma}$ the above equation takes the form:
\beq
- \rho {M \over r^2}  = (\Gamma-1) \rho^{\Gamma-1} {d\rho \over dr}.
\eeq
We can write this as:
\beq
- {M \over r^2} dr = (\Gamma-1) \rho^{\Gamma-2} d\rho.
\eeq

In large radii we can treat $M$ as constant and integrate both sides to get the following approximation:
\beq
{\rm M \over R}\propto ({\rm M \over \ \  R^3})^{\Gamma-1},
\eeq
which gives:
\beq
\rm M \propto R^{(3\Gamma-4) /( \Gamma-2)}.
\eeq

As $\Gamma=9/5$ here we will have $\rm {M \propto R^{-7}}$, which is the behaviour seen in large radii in Figure \ref{fig:logMRpoly}.


\section{Numerical solutions for realistic equations of state}

\label{sec:realistic}

\begin{figure}[t]
\centering
   \includegraphics[scale=0.4]{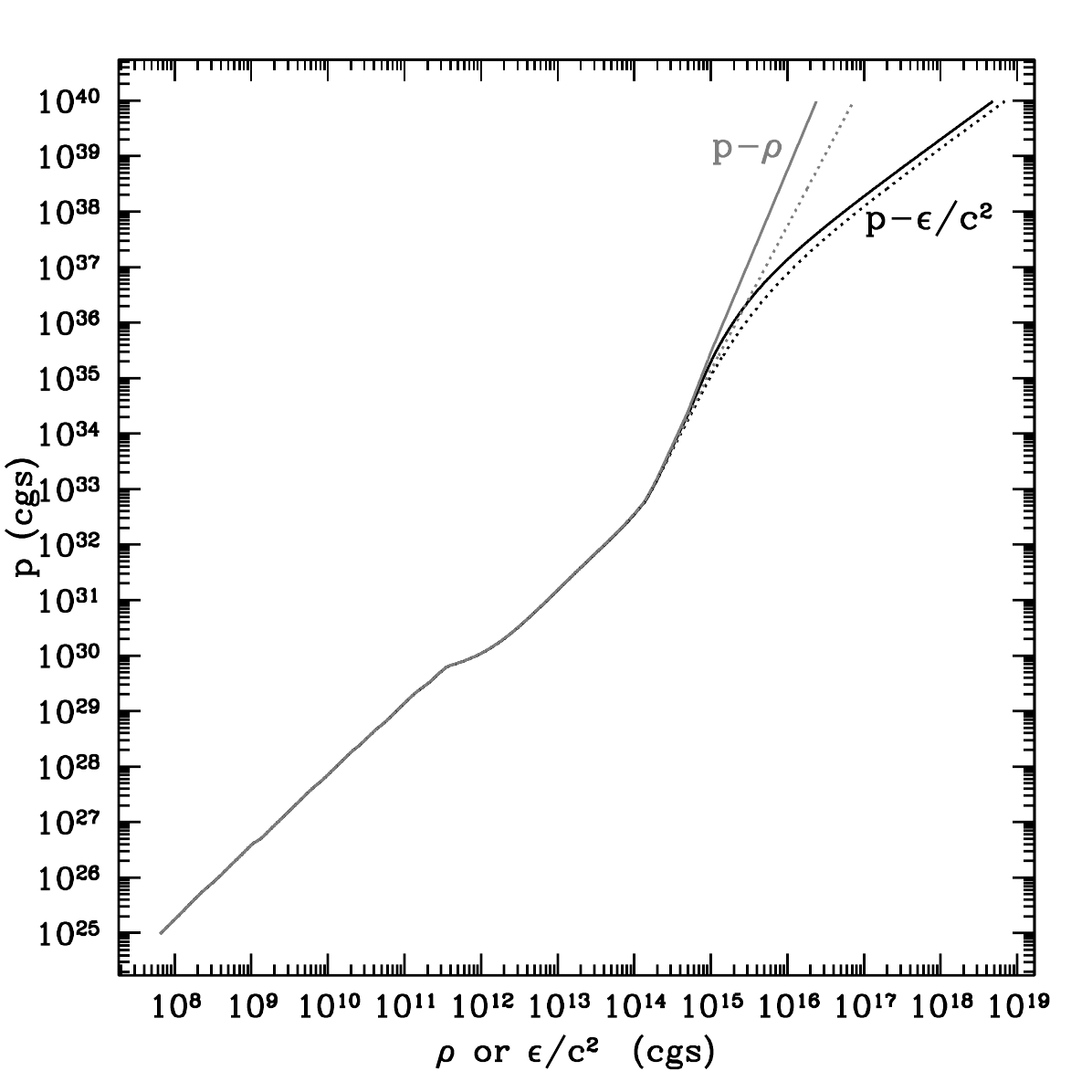}
\caption{$p-\rho$ (grey) and $p-\epsilon/c^2$ (black) relations: For matter densities $\rho<\rho_0=10^{14.3} {\rm g/cm^3}$ we use the SLy EOS \cite{douchin}. For $\rho>\rho_0$, the solid curves show the $p-\rho$ (grey) and $p-\epsilon/c^2$ (black) relations based on the minimal representation of the AP3 equation of state using the polytropic parameters of \cite{ozel}. The dotted curves are the same relations for the FPS equation of state. The equations have been smoothed to avoid discontinuities in the derivative of pressure as the aether theory is sensitive to these derivatives.}
\label{fig:EOS}
\end{figure}

\begin{figure}[t]
\centering
   \includegraphics[scale=0.4]{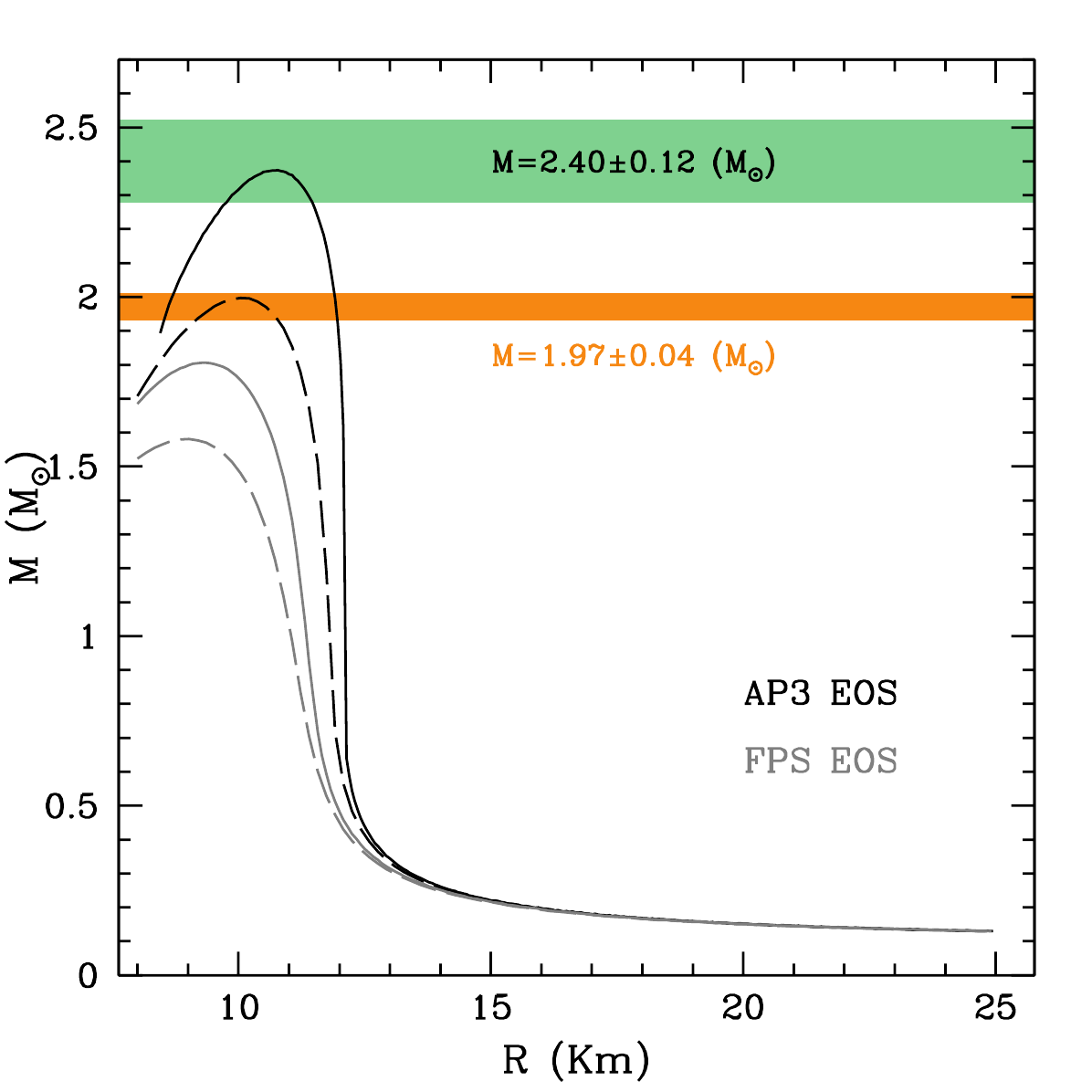}
\caption{(Color online) The ${\rm M}({\rm R})$ - R relation of neutron stars given by general relativity (solid) and the aether theory (dashed) based on the parametrized AP3 (black) and FPS (grey) equations of state. The two observed pulsar masses of Demorest et al.\ \cite{demorest} and van Kerkwijk et al.\ \cite{blackwidow} are shown in orange and green respectively.}
\label{fig:MREOS}
\end{figure}

For densities below a fiducial density $\rho_0$ of the order of the nuclear saturation density $\rho_{\rm ns} \sim 2.7 \times 10^{14} {\rm \ g/cm^3}$, the equation of state is well described by the SLy EOS \cite{douchin}. For densities higher than $\rho_0$, it is shown in \cite{ozel} that it is sufficient to specify the pressure of the neutron star at three fiducial densities $\rho_1=1.85 \rho_{\rm ns}$, $\rho_2=2 \rho_{1}$ and $\rho_3=2 \rho_{2}$ in order to reconstruct the mass-radius relations based on the AP3 and FPS equations of state. Figure \ref{fig:EOS} shows the SLy equation of state in $\rho<\rho_0$ as well as the parametrized AP3 and FPS equations of state in $\rho>\rho_0$ using the polytropic parameters found by \"Ozel and Psaltis and listed in table \ref{table:EOS} (from \cite{ozel}). 
\begin{table}[b]
\caption{\label{tab:table1}Polyropic parameters for the FPS and AP3 equations of state calculated by \"Ozel and Psaltis in \cite{ozel}.}
\begin{ruledtabular}
\begin{tabular}{lcccc}
EOS & $\log \rho_0$ & $\log P_1$ & $\log P_2$ & $\log P_3$ \\
\hline
FPS & 14.30 & 34.283 & 35.142 & 35.925 \\
AP3 & 14.30 & 34.392 & 35.464 & 36.452 \\
\end{tabular}
\end{ruledtabular}
\label{table:EOS}
\end{table}

If we use the smoothed version of these two equations of state (to avoid discontinuities in the derivative of pressure as the aether theory is sensitive to these derivatives) to numerically solve the stellar structure equations in the same way we did for a polytropic EOS in section \ref{sec:polytropic}, we find the ${\rm M}({\rm R})$ - R relations shown in Figure \ref{fig:MREOS}. General relativity predicts a maximum neutron star mass ${\rm M_{max}} \sim 1.81$ M$_\odot$ for the FPS EOS and ${\rm M_{max}} \sim 2.37$ M$_\odot$ for the AP3 EOS. Therefore the difference in the two parametrized equations of state (as seen by looking at  the difference of the solid and dotted curves in Figure \ref{fig:EOS}) results in a 24\% difference in the maximum neutron star mass predicted by general relativity. The aether theory gives a smaller maximum mass as expected from Section \ref{sec:polytropic}. The aether theory predicts ${\rm M_{max}} \sim 1.58$ M$_\odot$ for the FPS EOS and ${\rm M_{max}} \sim 2.00$ M$_\odot$ for the AP3 EOS. 

If the FPS equation of state is refuted on the basis of the 1.97 $\pm$ 0.04 M$_\odot$ neutron star observed by Demorest et al.\ \cite{demorest} and if we assume the validity of the AP3 equation of state, then both the aether theory and general relativity agree with this observational measurement. For this EOS the maximum mass predicted by the Aether theory is 16\% less than the maximum mass predicted by general relativity. However, for the AP3 EOS the aether prediction is inconsistent with the van Kerkwijk et al.\ \cite{blackwidow} measured mass, although this mass measurement is uncertain due to assumptions made in the theoretical model used in calculating the neutron star mass. As other EOS candidates exist (see \cite{lattimer2001}) more measurements of the mass and radius of neutron stars are needed to put further constraints on the existing equations of state, and allow us to make definite comparisons between the gravitational aether theory and general relativity.

As was mentioned in section \ref{cosmo}, the value of $\zeta_4$ for the aether theory is 1/3 in contrast with general relativity ($\zeta_4 =0$). Therefore, assuming the AP3 equation of state, the maximum neutron star masses given by the aether theory and general relativity in Figure \ref{fig:MREOS} can be translated into constraints on the value of $\zeta_4$. Using a linear interpolation, we find that ${\zeta_4}<0.43 \ (0.19)$ at 95\% confidence from the Demorest et al.\ \cite{demorest} (van Kerkwijk et al.\ \cite{blackwidow}) mass measurements.

\section{The aether equation of state}

\label{sec:aetherEOS} 

\begin{figure}[t]
\centering
   \includegraphics[scale=0.4]{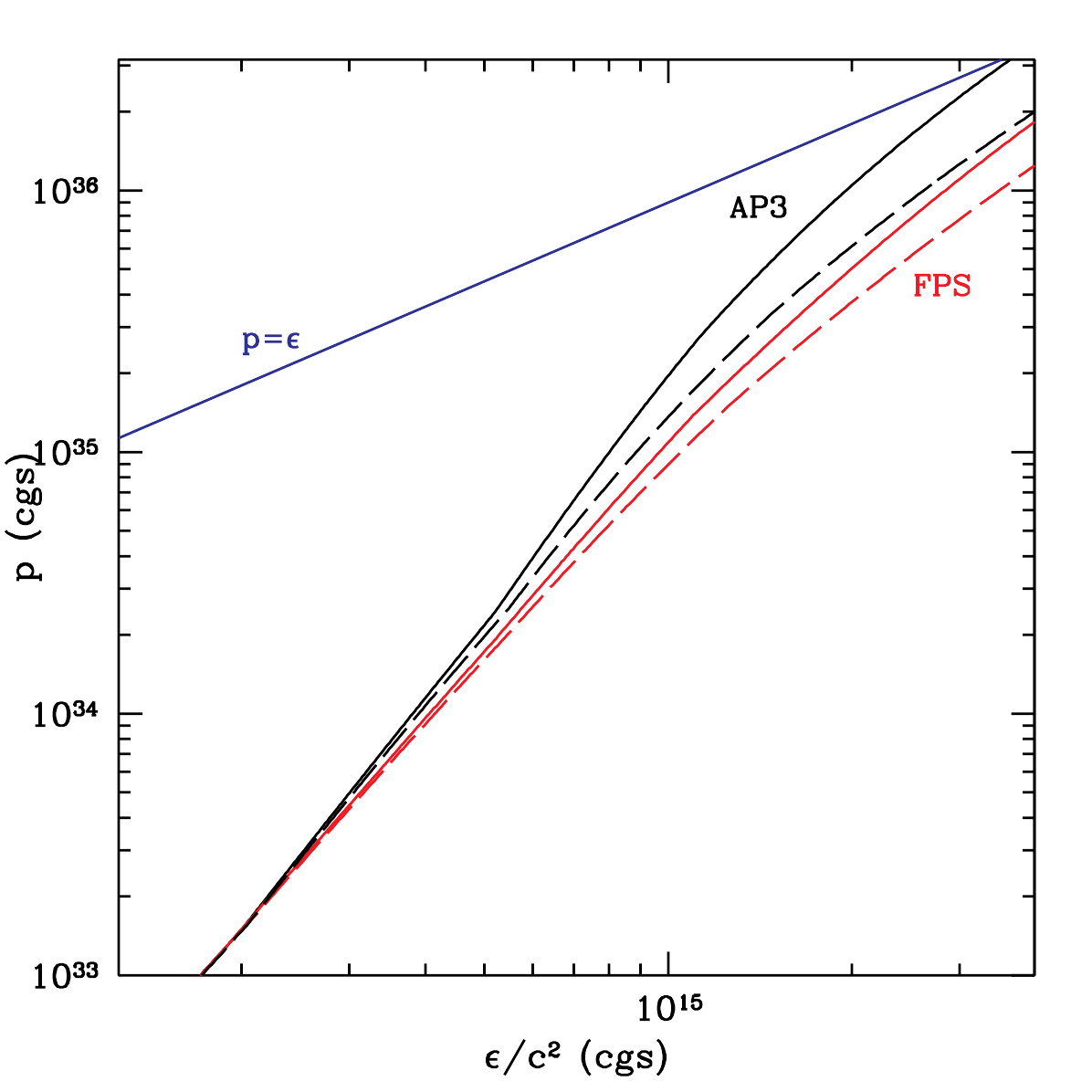}
\caption{(Color online) The AP3 (solid black) and FPS (solid red) parametric equations of state ${p}-{\epsilon}/c^2$ and the modified effective equations of state $\tilde{p}-\tilde{\epsilon}/c^2$ given by the gravitational aether theory based on the same equations of state (dashed black for AP3 and dashed red for FPS). The blue line corresponds to $p=\epsilon$.}
\label{fig:EOSAether}
\end{figure}

An equivalent description of the problem is suggested by Eq.\  \ref{densityandpressure}. This equation gives the updated energy density and pressure for which (with the new gravitational constant) the aether theory's Einstein equation (\ref{modein}) looks like the Einstein equation of general relativity. Therefore we can describe the aether theory's prediction of the structure of the neutron star as equivalent to the one of general relativity only with an updated equation of state given by $\tilde{p}(\tilde{\epsilon})$. To find this updated equation of state we equate the right-hand sides of equations \ref{eq3} and \ref{eq4} to get:
\beq
- {2 {p}'\over {\epsilon} + {p}}=- {2 \tilde{p}'\over \tilde{\epsilon} + \tilde{p}}.
\eeq

If we use Eq.\  \ref{densityandpressure} to write $\tilde{\epsilon}$ as a function of $\epsilon$ and $p$ and if we replace the derivatives with respect to radius with derivatives with respect to $\epsilon$ we get:
\beq
{d\tilde{p} \over d\epsilon} - {f(\epsilon) \over {\epsilon + p(\epsilon)}}\tilde{p} - f(\epsilon) =0,\label{diffeq}
\eeq
where $f(\epsilon)=dp(\epsilon)/d\epsilon$. Both $f(\epsilon)$ and $p(\epsilon)$ are given by the equation of state we are using. If we solve the differential equation \ref{diffeq} for the FPS and AP3 equations of state numerically we get $\tilde{p}$ as a function of $\epsilon$. Using Eq.\ \ref{densityandpressure} we can find $\tilde{p}$ as a function of $\tilde{\epsilon}$. These are the dashed curves shown in Figure \ref{fig:EOSAether}. The figure compares the AP3 and FPS parametric equations of state ($p-\epsilon$) and the modified equations of state ($\tilde{p}-\tilde{\epsilon}$) given by the gravitational aether theory based on the same equations of state. The effects of the aether theory become distinguishable beyond the nuclear saturation density (the region shown in Figure \ref{fig:EOSAether}). As in high densities, the aether theory gives a lower pressure than the one given by the equation of state, the stability of the neutron star will be obtained for a lower mass compared to the prediction of general relativity for a neutron star of the same radius. This is why the aether mass-radius relation falls below the general relativistic mass-radius relation for neutron stars.

\section{Conclusions and Future Prospects}

\label{sec:conclusion} 

The gravitational aether theory provides a possible solution to the cosmological constant problem. The structure of neutron stars is related to their nuclear properties (the equation of state) as well as the theory describing gravity. Therefore, the aether theory can be tested in the light of mass-radius measurements of these stars. In this paper, we solved the aether theory's equations of stellar structure for two equations of state of nuclear matter (AP3 and FPS) and found the mass-radius relation of neutron stars based on these two EOS, and compared this with the mass-radius relation given by general relativity. The FPS equation of state gave mass-radius relations that both for general relativity and the aether theory, were incompatible with the 1.97 $\pm$ 0.04 M$_\odot$ neutron star observed by Demorest et al.\ \cite{demorest}. The mass-radius relations given by the aether theory and general relativity on the basis of the AP3 equation of state were both compatible with this observation. We saw that for this equation of state the aether predicts a maximum mass that is 16\% less than the maximum mass predicted by general relativity. We also found the modified equation of state of neutron stars given by the aether theory and based on that explained why the mass-radius relation given by the aether theory falls below the one given by general relativity. It is important to note that there are other equations of state such as the one calculated by  M\"uller et al.\ \cite{muller} (and others mentioned in \cite{lattimer2001}) that do also agree with the Demorest et al.\ measured pulsar mass. 

In addition, including the effect of hyperons and quarks in the equation of state (e.g.\ \cite{alford, prakashhyperon, glendenning85, balberg, glendenning99, pandharipande75}) can have a similar effect to the aether in lowering the maximum mass of neutron stars. For example, it is shown in \cite{alford} that a hybrid (nuclear+quark matter) star can have a mass-radius relationship very similar to that predicted for a star made of purely nucleonic matter. The authors obtain hybrid stars as heavy as 2 M$_\odot$ for reasonable values of their model parameters. Due to these uncertainties in the equation of state we can not make definite comparisons between the aether theory and general relativity at the moment.

To be able to test the aether theory more robustly, we need further constraints on the neutron star equations of state. In addition to constraints coming from the mass measurements of neutron stars (such as \cite{demorest}), we also need further constraints from radius measurements that are considerably harder to get. The radius of a neutron star can be measured in various ways such as the thermal observations of the surface of the neutron star, pulse profiles or redshift measurements \cite{lattimer2007}. We also know that mergers of compact objects such as pairs of neutron stars or neutron star-black hole pairs emit gravitational waves. These waves  can be detected using current detectors if the emitter is close enough. Gravitational waves allow us to simultaneously measure masses and radii of these compact objects and could constrain the neutron star maximum mass and thus its equation of state \cite{lattimer2007}. They can also constrain the equation of state directly. In \cite{gwave}, the authors have studied the accuracy with which one can use gravitational wave observations of double neutron star inspirals to measure parameters of the neutron-star equation of state using numerical simulations. They concluded that gravitational wave observations could put a direct constraint on the EOS pressure at a rest mass density $\rho = 5 \times 10^{14}$ $\rm g \ cm^{-3}$ of $\delta p \sim 10^{32} \ \rm dyn \ cm^{-2}$ at an effective distance $D_{\rm eff}=100$ Mpc (also see \cite{amsterdam}). At this density, the difference of pressure between the aether EOS and the AP3 EOS is $\Delta p = 7.6 \times 10^{32} \ \rm dyn \ cm^{-2}$. This means that the aether theory's modified equation of state can be tested using gravitational waves, unless its predicted pressure value at the density mentioned above, is equal to the pressure predicted by another equation of state. This degeneracy will fade if further progress is made in constraining the equation of state of nuclear matter in densities above the nuclear saturation density. Another promising way to break this degeneracy is to study the dynamics of a collapsing neutron star which could distinguish the effects of modifying gravity from modifying the equation of state.

Future observations ranging from the electromagnetic emissions of pulsars to the gravitational wave emissions of neutron stars in compact systems will allow us to learn not only about the nuclear constituents at the core of neutron stars but also about the nature of gravitation and fundamental questions such as the cosmological constant problem.

\vspace*{10mm}

We would like to acknowledge the helpful discussions with Dimitrios Psaltis and also thank him for providing us with data for various equations of state. We are also grateful to Latham Boyle, Luis Lehner and Rafael Sorkin for their insights. We are supported for this research by the University of Waterloo and the Perimeter Institute for Theoretical Physics. Research at Perimeter Institute is supported by the Government of Canada through Industry Canada and by the Province of Ontario through the Ministry of Research \& Innovation.


\bibliographystyle{h-physrev5}

\bibliography{Neutron_Star_Aether}

\end{document}